\documentclass[twocolumn,amsmath,amssymb,prl,superscriptaddress,nobibnotes,showpacs]{revtex4}
\usepackage{graphicx}

\begin{document}

\title{Preparation and characterization of arbitrary states of four-dimensional qudits based on biphotons}

\author{So-Young Baek} \email{simply@postech.ac.kr}
\affiliation{Department of Physics, Pohang University of Science and Technology (POSTECH), Pohang, 790-784, Korea}

\author{Stanislav S. Straupe}
\affiliation{Department of Physics, Moscow State University, Moscow, 119899, Russia}

\author{Alexander P. Shurupov}
\affiliation{Department of Physics, Moscow State University, Moscow, 119899, Russia}

\author{Sergei P. Kulik}
\affiliation{Department of Physics, Moscow State University, Moscow, 119899, Russia}

\author{Yoon-Ho Kim}
\email{yoonho@postech.ac.kr}
\affiliation{Department of Physics, Pohang University of Science and Technology (POSTECH), Pohang, 790-784, Korea}


\date{\today}

\begin{abstract}
We report interferometric schemes to prepare arbitrary states of
four-dimensional qudits (ququarts) based on biphoton states of
ultrafast-pumped frequency-nondegenerate spontaneous parametric
down-conversion. Preparation and tomographic characterization of
a few examples of general single-ququart states, a pure state,
a completely mixed state, and a partially-mixed state, are experimentally
demonstrated.
\end{abstract}

\maketitle


\section{Introduction}

In quantum information, a two-dimensional quantum system is often used as a carrier of the basic information unit, the two-level quantum state or the quantum bit (qubit).  Most quantum computing and quantum communication protocols are based on preparation, manipulation, entanglement, distribution, and measurement of multiple qubits \cite{chuang}.

Recently, $D$-level quantum states ($D>2$) or qudits have attracted a lot of attention in the context of quantum communication and experimental tests of quantum mechanics \cite{rungta,bouda,peres,bech,boure,kasz,collins,fu}. Experimentally, the physical carrier of the qudit can be any $D$-dimensional quantum systems. Meaningful applications of qudits in quantum information, however, will only be possible if the qudit is encoded in a $D$-dimensional physical degree of freedom which is easy to handle experimentally. Furthermore, it should be possible to entangle multiple individual qudits in a scalable manner.

In photonic quantum information research, an internal or an external degree of freedom of a photon is used to encode the intended quantum state. For a qubit, the polarization state of a photon is often the obvious choice although it is possible to choose other degrees of freedom. To encode a qudit, it is necessary to choose a multi-dimesional degree of freedom of a single-photon, such as, the angular momentum, transverse momentum-position, time of arrival, etc \cite{mair,vaziri,ried,thew04,langford,hale,neves}. These single-photon multi-dimensional photonic degrees of freedom, however, are experimentally difficult to manipulate and there are no scalable schemes to generate multi-qudit entangled states.

The single qudit, however, does not need to be encoded in a single-particle quantum state. In fact, preparation and tomographic characterization of a pure state qutrit (three-dimensional quantum state), see Refs.~\cite{burlakov,bogdanov1,bogdanov2}, and a pure state ququart (four-dimensional quantum state), see Refs.~\cite{moreva1,bogdanov3}, have been demonstrated recently using the biphoton polarization states of frequency-degenerate and frequency-nondegenerate spontaneous parametric down-conversion, respectively. In other words, a pair of photons can be used as a carrier of three- or four-dimensional quantum states.

Especially, the ququart based on the biphoton polarization state of frequency-nondegenerate spontaneous parametric down-conversion (SPDC) exhibits a few properties which are important for applications in quantum information research: First, all the ququart basis states can be accessed using only linear optical elements (phase plates) \cite{moreva1,bogdanov3}. Second, it is possible to prepare a multi-ququart entangled state starting from multiple individual biphoton ququarts, linear optical elements (beam splitters), and post-selection measurement \cite{baek1,baek2}. So far, only pure state biphoton ququarts have been demonstrated. It is, therefore, of interest and importance to learn how to prepare a general state of a biphoton ququart \cite{disadvantage}.

In this paper, we report experimental studies on preparation of
general states of biphoton ququarts using ultrafast-pumped
frequency-nondegenerate spontaneous parametric down-conversion.
Methods for preparation and tomographic characterization of some
examples of arbitrary ququart states, i.e., pure, mixed, and
partially-mixed states of a biphoton ququart, are experimentally
demonstrated. We also discuss a couple of alternative experimental
schemes which allow to generate arbitrary biphoton-based ququart
states.

\section{Biphoton ququart}

Let us start with a brief introduction to the biphoton ququart. In collinear frequency-nondegenrate SPDC, a higher energy pump photon is occasionally split into a pair of co-propagating lower energy photons (signal-idler) of different frequencies,
\begin{equation}
\frac{1}{\lambda_p} = \frac{1}{\lambda_1}+\frac{1}{\lambda_2},
\end{equation}
where $\lambda_p$ is the pump wavelength and $\lambda_1$ ($\lambda_2)$ is the wavelength of the signal (idler) photon. Since each photon of the pair can be horizontally or vertically polarized, the following set of biphoton polarization basis states can be defined \cite{moreva1,bogdanov3,baek1,baek2},
\begin{eqnarray}\label{basis}
|H_{\lambda_1},H_{\lambda_2}\rangle \equiv |0\rangle,~~ 
|H_{\lambda_1},V_{\lambda_2}\rangle \equiv |1\rangle, \nonumber\\
|V_{\lambda_1},H_{\lambda_2}\rangle \equiv |2\rangle,~~ 
|V_{\lambda_1},V_{\lambda_2}\rangle \equiv |3\rangle.\nonumber
\end{eqnarray}
The polarization state of the photon pair born in the process of collinear frequency-nondegenerate SPDC, therefore, represents a four-dimensional quantum state or a ququart. Since the states shown in eq.~(\ref{basis}) are orthonormal to each other and form a complete basis for a four-dimensional Hilbert space, they form the computational basis for the biphoton ququart.

Experimentally, the computational basis states for the biphoton ququart can be generated using type-I SPDC ($|0\rangle$ and $|3\rangle$) and type-II SPDC ($|1\rangle$ and $|2\rangle$). To prepare an arbitrary superposition state of a ququart,
\begin{equation}\label{superposition}
|\psi\rangle = c_0 |0\rangle + c_1 |1\rangle + c_2 |2\rangle + c_3 |3\rangle,
\end{equation}
where $c_l=|c_l| e^{i \phi_l}$ are the complex probability amplitudes which satisfy $\sum_{l=0}^{3}|c_l|^2=1$, one would need to coherently combine two type-I SPDC and two type-II SPDC sources and to control four complex amplitudes $c_l, c_2, c_3$, and $c_4$. The use of four SPDC sources, however, turns out to be unnecessary since the degree of polarization of the biphoton state of collinear frequency-nondegenerate SPDC is not invariant under SU(2) transformations \cite{karassiov}. It is, therefore, possible to use linear optical elements (wave plates) to transform one biphoton ququart basis state into any other one as recently demonstrated in Ref.~\cite{moreva1,bogdanov3}. As a result, it is possible to prepare an arbitrary superposition state (i.e., pure state) shown in eq.~(\ref{superposition}) using less than four SPDC sources and we discuss several such schemes in this paper.

The superposition state shown in eq.~(\ref{superposition}), however, is not the most general quantum state for a ququart. To properly consider mixedness, which comes from controlled or uncontrolled quantum distinguishability among the basis states shown in eq.~(\ref{basis}), the ququart state should be expressed as a $4\times4$ density matrix $\rho$ to describe general single-ququart states: completely mixed states, partially-mixed states, and the pure state shown in eq.~(\ref{superposition}).

It is interesting to  note that, since the ququart state under consideration is in fact formed with two polarization qubits, it is possible to apply the state classification method based on two-qubit entanglement of formation. This method is based on the quantity $C$, concurrence, which is a measure of two-qubit entanglement \cite{wooters}. For the single biphoton ququart (i.e., biphoton two-qubit state) in eq. (\ref{superposition}), it is easy to show that there is no two-qubit entanglement ($C = 0$, i.e., the two-qubit state is separable) if the amplitudes satisfy the relation $c_0 c_3 = c_1 c_2$. Otherwise, the state is an entangled two-qubit state with concurrence  $C = 2|c_0 c_3 - c_1 c_2| > 0$. (The two-qubit state is non-maximally entangled if $0<C<1$.)

\section{Ququart preparation}

In order to prepare a general single-ququart state, including mixed, partially-mixed, and pure states, it is necessary to introduce quantum distinguishability among the biphoton ququart basis states defined in eq.~(\ref{basis}) in a controllable manner. Moreover, to establish a confidence bound and to find a reliable method of preparing an intended ququart state, it is required to compare the experimentally reconstructed and the theoretically expected ququart density matrices. Our experiment, therefore, deals with both the ququart preparation as well as the experimental reconstruction of the ququart density matrices using the quantum state tomography.

\begin{figure}[t]
\centering\includegraphics[width=3.4in]{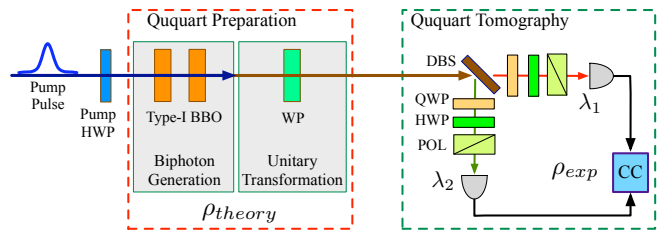}
\caption{Schematic of the experiment. Ququart preparation: The collinear frequency-nondegenerate biphoton polarization state is transformed by the wave plate (WP) to prepare an arbitrary four-dimensional quantum state, a biphoton ququart. The theoretically expected quantum state of the ququart, $\rho_{theory}$, can be calculated by using the information on the experimental settings. Ququart tomography: The biphoton ququart is tomographically characterized to obtain the experimentally reconstructed density matrix $\rho_{exp}$. See text for details.}
\label{setup1}
\end{figure}

The schematic of the experimental setup is shown in Fig.~\ref{setup1}. In the single-crystal scheme, a type-I BBO crystal, with its optic axis horizontally oriented, is pumped by a train of ultrafast pulses and this scheme is used for preparing pure state ququarts. For preparing mixed and partially-mixed states, we used the double-crystal scheme in which two orthogonally oriented type-I BBO crystals are placed in tandem \cite{kim1}. The pump laser had roughly 100 fs pulse width and centered at 390 nm. The polarization of the pump was controlled by using a half-wave plate (Pump HWP). The BBO crystals used on this experiment are 3 mm thick. The SPDC signal-idler photon pair generated at the crystals propagates collinearly with the pump and has different frequencies: the signal photon is centered at 823.5 nm ($\lambda_1$) and the idler photon  is centered at 740.8 nm ($\lambda_2$).

In the single-crystal scheme, the biphoton polarization state generated at the crystal is  $|V_{\lambda_{1}},V_{\lambda_{2}}\rangle$ which corresponds to the ququart basis state $|3\rangle$ in eq.~(\ref{basis}). This basis state is then unitarily transformed into a superposition state shown in eq.~(\ref{superposition}) to prepare an arbitrary pure state by using a zero-order wave plate (WP).  This is due to the fact that the signal and the idler photons, even though they have the same polarization, acquire different phase shift due to the large difference in the wavelength. As a result, the polarization states of the signal and the idler photons evolve differently, but predictably, for a specific angle of WP: $|V_{\lambda_1}\rangle \rightarrow \alpha |H_{\lambda_1}\rangle + \beta |V_{\lambda_1}\rangle$ and $|V_{\lambda_2}\rangle \rightarrow \gamma |H_{\lambda_2}\rangle + \delta |V_{\lambda_2}\rangle$ with $|\alpha|^2 + |\beta|^2 = 1$ and $|\gamma|^2 + |\delta|^2 = 1$. The final result is the transformation of the initial ququart basis state into a superposition state in eq.~(\ref{superposition}) which is a pure state. The theoretically expected density matrix $\rho_{theory}$ for the prepared ququart state, which can be calculated accurately as the settings of WP is in our control, is then compared to the experimentally reconstructed density matrix $\rho_{exp}$.

The single-crystal scheme, therefore, allows us to prepare one of the four single-ququart basis states in eq.~(\ref{basis}) initially. For the case just described, the single-ququart basis state $|3\rangle$ is in fact $|V_{\lambda_{1}},V_{\lambda_{2}}\rangle$, which is a factorizable two-qubit state with $C=0$. Note now that the degree of entanglement (in this case, between the two polarization qubits) cannot be increased or decreased by local unitary transformations. The unitary transformation due to WP can turn the initial ququart state $|3\rangle$ in to a superposition state in the form of eq.~(\ref{superposition}). However, the resulting state can always be expressed as a separable two-qubit state of the form $|A_{\lambda_{1}},B_{\lambda_{2}}\rangle$, where $|A_{\lambda_{1}}\rangle$ and $|B_{\lambda_{2}}\rangle$ represent arbitrary polarization states of signal and idler photons, respectively.

For preparation of mixed and partially-mixed ququart states, the double-crystal scheme described earlier is used.   Since the two BBO crystals are orthogonally oriented, it is possible to excite two of the ququart basis states shown in eq.~(\ref{basis}): $|0\rangle$ ($|H_{\lambda_{1}},H_{\lambda_{2}}\rangle$) and $|3\rangle$ ($|V_{\lambda_{1}},V_{\lambda_{2}}\rangle$). The relative amplitudes between the two can be controlled by changing the pump polarization.

To prepare a completely mixed state of $|0\rangle$ and $|3\rangle$, no further actions are required as the two amplitudes are already distinguishable in time due to the clock effect of the pump pulse \cite{kim1}. The resulting density matrix is, therefore,
\begin{equation} \label{mix}
\rho_{theory}= \left(1-\frac{x}{2}\right)|0\rangle\langle0| +\frac{x}{2}|3\rangle\langle3|,
\end{equation}
where the parameter $x$ can be varied by changing the polarization of the pump. Adding two additional type-II BBO crystals, which are orthogonally oriented, will allow us to easily prepare a mixture of all four ququart basis states. This, however, is not necessary in principle since it is possible to transform a single basis state into a superposition of all basis states and then to introduce birefringent/dichroic decoherence among the amplitudes.

A more general state, between the completely mixed and the pure states, would exhibit some coherence among the four ququart basis states in eq.~(\ref{basis}). In other words, the ququart density matrix has non-zero off-diagonal elements. Such states can be prepared by unitarily transforming the mixed
state in eq.~(\ref{mix}) using WP. As previously discussed, a
ququart basis state can be transformed into a superposition of all
basis states linear optically using WP \cite{karassiov}. By
subjecting the mixed state in eq.~(\ref{mix}) to unitary
transformation using WP, it is possible to obtain the following
partially-mixed state,
\begin{equation} \label{int}
\rho_{theory}= p_1 |\psi_1\rangle\langle\psi_1| +  p_2 |\psi_2\rangle\langle\psi_2|,
\end{equation}
where $p_1+p_2=1$ and $|\psi\rangle_1$, for example, is in the form of eq.~(\ref{superposition}). It is important to note that this unitary transformation process, however, does not actually decrease entropy of the ququart state as we shall show in the next section. To decrease the entropy, it is necessary to erase the temporal distinguishability of the biphoton amplitudes born in the first and the second crystals, for example, by inserting a piece of thick compensating quartz crystal in the pump beam or in the path of the photon pair \cite{kim1}. Introduction of controllable birefringent/dichroic decoherence will transform eq.~(\ref{int}) into a more complex ququart state with increased entropy.

From the two-qubit perspective, linear optical state transformation
from eq.~(\ref{mix}) to eq.~(\ref{int}) represents no increase in
the degree of two-qubit entanglement as both states exhibit the
two-qubit concurrence $C=0$. This is closely related to the fact
that the ququart state entropy remains the same for states in
eq.~(\ref{mix}) and eq.~(\ref{int}). To be more specific, the
two-qubit concurrence $C$ will increase if the single-ququart
entropy is decreased by erasing the temporal distinguishing
information present in eq.~(\ref{mix}) or in eq.~(\ref{int}). Complete erasure of the temporal distinguishability between the biphoton amplitudes from the first ($|H_{\lambda_{1}},H_{\lambda_{2}}\rangle$) and the second ($|V_{\lambda_{1}},V_{\lambda_{2}}\rangle$) type-I BBO crystal, see Fig.~\ref{setup1},  will result a pure ququart state with the two-qubit concurrence $C=1$.

In this paper, we have experimentally demonstrated a pure, a mixed, and a partially mixed biphoton ququart states which are shown in  eq.~(\ref{superposition}), eq.~(\ref{mix}), and eq.~(\ref{int}), respectively. For the mixed and partially mixed states, the state entropy can be controlled by inserting a proper compensating crystal before or after the BBO crystals.

\section{Ququart tomography}

The prepared ququart state is characterized experimentally by performing quantum state tomography, a
statistical method of reconstructing the quantum state density matrix based on a set of polarization projection measurement \cite{james,thew}.

The experimental schematic for the ququart state tomography is shown
in Fig.~\ref{setup1}. First, the photon pair that forms the biphoton
ququart is split into two spatial modes by using a dichroic beam
splitter (DBS), which transmits $\lambda_1 = 823.5$ nm and reflects
$\lambda_2 = 740.8$ nm. Each photon, then, undergoes polarization
state transformation with the use of a quarter wave plate (QWP) and
a half wave plate (HWP). Finally, the polarization state projection
is applied to each photon by using a polarizer (POL) after which the
photon is detected at the detector package, which consists of a
spectral filter and a multi-mode fiber coupled single-photon
counting module (SPCM). Since the ququart is made of a pair of
photons, the ququart detection is based on the coincidence counting
rates of the two SPCM's that detects $\lambda_1$ and $\lambda_2$
photons with definite polarizations. The coincidence window used in
this experiment was 5 ns.

As noted in Ref.~\cite{moreva1}, the ququart based on the collinear
frequency-nondegenerate SPDC photon pair is mathematically
equivalent to the non-collinear frequency-degenerate SPDC photon
pair which is often used in quantum information research. It is thus
possible to apply the two-qubit quantum tomography method described
in Ref.~\cite{james} directly in this experiment to reconstruct the
single-ququart density matrix. Sixteen particular joint biphoton
polarization state measurements are, therefore, necessary to
reconstruct a single-ququart density matrix for the identically
prepared ensemble of biphoton ququarts. Table \ref{tb} shows
experimental settings of WP to perform the sixteen polarization
projection measurements. The set of sixteen coincidence measurement
outcomes $n_{\nu}$ allow the linear tomographic reconstruction of
the single-ququart density matrix. It is, however, possible that the
mathematically reconstructed density matrix by the linear
tomographic reconstruction might violate the physical properties of
a density matrix. To avoid this problem, the maximum likelihood
method was applied as follows \cite{james}. First, we generate a
physical density matrix which satisfies normalization, Hermiticity,
and positivity, as a function of sixteen variables. We then
introduce the ``likelihood function" which quantifies how good
physical density matrix is in relation to the experimental data.
Finally, using standard numerical optimization techniques, we obtain
the best estimate of the density matrix by maximizing the likelihood
function. The single-ququart density matrix initially reconstructed
by the linear tomography is used as the seed for the iteration
algorithm.

The experimentally reconstructed ququart density matrix $\rho_{exp}$
is then compared to the theoretically expected density matrix
$\rho_{theory}$ which is calculated from the known values of the
pump polarization and WP settings. The fidelity
$F=(Tr\sqrt{\sqrt{\rho_{theory}}\rho_{exp}\sqrt{\rho_{theory}}})^2$
is then calculated to see how closely the two overlap and the state
purity is analyzed by calculating the state entropy defined as $S =
- \sum_{k=1}^{4} \lambda_k \log_4 \lambda_k$, where $\lambda_k$ are
the eigenvalues of the density matrix $\rho$.

\begin{table}[t]
\centering
  \caption{QWP and HWP (fast axis) settings for ququart tomography. POL in Fig.~\ref{setup1} transmits vertical polarization. Here, $|D\rangle=(|H\rangle+|V\rangle)/\sqrt{2}$, $|A\rangle=(|H\rangle-|V\rangle)/\sqrt{2}$,
  $|D\rangle=(|H\rangle+|V\rangle)/\sqrt{2}$, and $|R\rangle=(|H\rangle+i|V\rangle)/\sqrt{2}$.
}
\begin{tabular}{c c c c c c c}
  \hline \hline
  $\nu$ & HWP1 & QWP1 & HWP2 & QWP2 & basis1 & basis2 \\[0.5ex]\hline
  1 & $45^{\circ}$ & $0$ & $45^{\circ}$ & $0$  & $|H\rangle$ & $|H\rangle$  \\
  2 & $45^{\circ}$ & $0$ & $0$ & $0$  & $|H\rangle$ & $|V\rangle$  \\
  3 & $0$ & $0$ & $0$ & $0$ & $|V\rangle$ & $|V\rangle$  \\
  4 & $0$ & $0$ & $45^{\circ}$& $0$  & $|V\rangle$ & $|H\rangle$  \\
  5 & $22.5^{\circ}$ & $0$ & $45^{\circ}$ & $0$  & $|R\rangle$ & $|H\rangle$  \\
  6 & $22.5^{\circ}$ & $0$ & $0$ & $0$  & $|R\rangle$ & $|V\rangle$  \\
  7 & $22.5^{\circ}$ & $45^{\circ}$ & $0$ & $0$  & $|D\rangle$ & $|V\rangle$  \\
  8 & $22.5^{\circ}$ & $45^{\circ}$ & $45^{\circ}$ & $0$  & $|D\rangle$ & $|H\rangle$  \\
  9 & $22.5^{\circ}$ & $45^{\circ}$ & $22.5^{\circ}$ & $0$  & $|D\rangle$ & $|R\rangle$  \\
  10 & $22.5^{\circ}$ & $45^{\circ}$ & $22.5^{\circ}$ & $45^{\circ}$  & $|D\rangle$ & $|D\rangle$  \\
  11 & $22.5^{\circ}$ & $0$ & $22.5^{\circ}$ & $45^{\circ}$  & $|R\rangle$ & $|D\rangle$  \\
  12 & $45^{\circ}$ & $0$ & $22.5^{\circ}$ & $45^{\circ}$  & $|H\rangle$ & $|D\rangle$  \\
  13 & $0$ & $0$ & $22.5^{\circ}$ & $45^{\circ}$  & $|V\rangle$ & $|D\rangle$  \\
  14 & $0$ & $0$ & $22.5^{\circ}$ & $90^{\circ}$  & $|V\rangle$ & $|L\rangle$  \\
  15 & $45^{\circ}$ & $0$ & $22.5^{\circ}$ & $90^{\circ}$  & $|H\rangle$ & $|L\rangle$  \\
  16 & $22.5^{\circ}$ & $0$ & $22.5^{\circ}$ & $90^{\circ}$  & $|R\rangle$ & $|L\rangle$
  \\[1ex]
  \hline
\end{tabular} \label{tb}
\end{table}

\subsection{Pure state ququart}

To prepare a pure state ququart as in eq.~(\ref{superposition}), we used a single type-I BBO crystal generating the initial ququart state $|3\rangle$, which in fact is a factorizable two-qubit state $|V_{\lambda_{1}},V_{\lambda_{2}}\rangle$. This state is then transformed into a superposition state with the help of WP shown in Fig.~\ref{setup1}. In this experiment, a zero-order half wave plate designed at 823.5 nm was used in place of WP. As a demonstration of pure state ququart preparation, we set the fast axis of the half-wave plate at $30^\circ$ from the vertical axis.  Since local unitary transformations do not change the degree of entanglement, the final states belong to the factorizable subset of two-qubit (ququart) states with $C=0$.

The theoretically expected ququart density matrix in this case is calculated to be,
\begin{widetext}
\begin{equation}
\rho_{theory}^{pure}=
    \left(
      \begin{array}{cccc}
        0.5432     &       0.3136+0.1182i      &       0.3136     &       0.1811+0.0683i \\
        0.3136-0.1182i     &       0.2068      &       0.1811-0.0683i      &       0.1194 \\
        0.3136     &       0.1811+0.0683i       &       0.1811     &       0.1045+0.0394i \\
        0.1811-0.0683i    &       0.1194       &       0.1045-0.0394i      &       0.0689\\
      \end{array}
    \right).
\end{equation}
\end{widetext}
For this state, it is easy to see that $Tr[\rho^2]=1$. The entropy
of the theoretical density matrix $\rho_{theory}^{pure}$ is
calculated to be $S = 0$, as it should for a pure state.

To obtain the ququart density matrix, we have performed the sixteen projection measurement described in Table \ref{tb}. The coincidence data (for the accumulation time of 180 s) are $n_{1}=6118$,
$n_{2}=1858$, $n_{3}=917$, $n_{4}=2943$, $n_{5}=1565$, $n_{6}=477$,
$n_{7}=2362$, $n_{8}=7549$, $n_{9}=2395$, $n_{10}=8254$,
$n_{11}=1664$, $n_{12}=6653$, $n_{13}=3078$, $n_{14}=2739$,
$n_{15}=5817$, and $n_{16}=1398$. The accidental coincidences, which appear at the same period as the pump pulse period, have been subtracted. In this paper, all the coincidence measurement data show accidental subtracted values.

\begin{figure}[t]
\centering\includegraphics[width=3.4in]{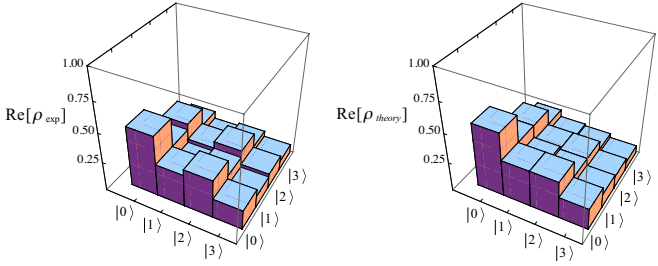}
\caption{Real part of the experimentally reconstructed ($\rho_{exp}$) and the theoretically expected ($\rho_{theory}$) density matrices for a pure state ququart. The calculated fidelity is $F=0.938$.}
\label{pure}
\end{figure}

Applying the single-ququart tomography algorithm, the experimentally
reconstructed ququart density matrix is obtained to be,
\begin{widetext}
\begin{equation}
\rho_{exp}^{pure}=
    \left(
      \begin{array}{cccc}
        0.5138     &       0.2749+0.0523i       &       0.3236+0.1308i      &       0.1643+0.1026i \\
        0.2749-0.0523i     &       0.1590       &       0.1887+0.0418i      &       0.1004+0.0379i \\
        0.3236-0.1308i     &       0.1887-0.0418i       &       0.2463      &       0.1259+0.0224i \\
        0.1643-0.1026i     &       0.1004-0.0379i       &       0.1259-0.0224i      &       0.0777\\
      \end{array}
    \right).
\end{equation}
\end{widetext}
We obtain $Tr[\rho_{exp}^2]=0.962$ which means that the experimentally reconstructed state is not an entirely pure state. The entropy of the experimentally reconstructed density matrix is found to be $S=0.052$.

Clearly, the experimentally prepared ququart state is somewhat
different from what we initially intended to prepare and this is
reflected in the state fidelity $F=0.938$. Figure \ref{pure} shows
the graphical representations of the real parts of the experimental
and theoretical ququart density matrices for  this experiment.

The errors in the experimentally reconstructed density matrix can be estimated as follows. Assuming that the measured coincidences $n_{\nu}$ follow Poissonian statistics, i.e., the
ensemble average of the uncertainties satisfy $\overline{\delta
n_{\nu}}=\sqrt{n_{\nu}}$, the error in the reconstructed density matrix is calculated to be,
\begin{widetext}
\begin{equation}
\Delta\rho_{exp}^{pure}=
    \left(
      \begin{array}{cccc}
        0.0066          &       0.0042-0.0034 i       &       0.0069+0.0023i      &       0.0083+0.0028i \\
        0.0042+0.0034i    &      0.0036       &       0.0056-0.0035i      &       0.0039+0.0013i \\
        0.0069-0.0023i     &       0.0056+0.0035i       &      0.0046      &       0.0029-0.0024i \\
        0.0083-0.0028i     &      0.0039-0.0013i       &       0.0029+0.0024i      &       0.0026\\
      \end{array}
    \right).
\end{equation}
\end{widetext}

\subsection{Mixed state ququart}

The double-crystal scheme shown in Fig.~\ref{setup1} is used to prepare a ququart in a completely mixed state. To study different degrees of mixedness for a biphoton ququart, we studied two cases: the ququart prepared with the pump polarization of $30^\circ$ and of $45^\circ$ from the horizontal plane.

Let us first discuss the case of $30^\circ$ pump polarization. In this case, the horizontally oriented type-I BBO is pumped more strongly than the vertically oriented type-I BBO. This setting, therefore, produces an unequal mixture of the ququart basis states $|3\rangle$ and $|0\rangle$.

\begin{figure}[t]
\centering\includegraphics[width=3.4in]{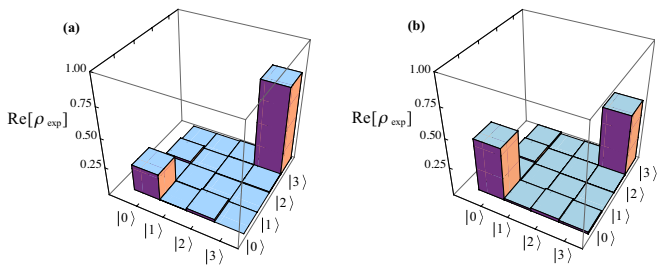}
\caption{Real parts of two completely mixed state ququarts with different entropy values. These states are generated using the double-crystal scheme shown in Fig.~\ref{setup1} with different pump polarization. (a) For the state shown in eq.~(\ref{mix1}). $S(\rho_{exp})=0.394$. (b) For the state shown in  eq.~(\ref{mix2}). $S(\rho_{exp})=0.504$.}
\label{mixed}
\end{figure}

The theoretical ququart density matrix for this case, as in eq.~(\ref{mix}), takes a very simple form,
\begin{equation}
\rho_{theory}^{mix}=
    \left(
      \begin{array}{cccc}
        0.2500     &       0     &      0     &       0 \\
        0     &       0      &       0      &       0 \\
        0     &       0       &       0     &       0 \\
        0    &       0       &       0      &       0.7500\\
      \end{array}
    \right).
\end{equation}
The theoretical density matrix is clearly not pure since
$Tr[\rho_{theory}^2] = 0.625$ and $S = 0.406$.

For this state, the tomographic coincidence measurement resulted (for $180$ s) $n_{1}=1911$, $n_{2}=46$, $n_{3}=6287$, $n_{4}=34$, $n_{5}=795$, $n_{6}=3562$, $n_{7}=3005$, $n_{8}=1048$,
$n_{9}=1911$, $n_{10}=2061$, $n_{11}=2321$, $n_{12}=981$,
$n_{13}=3141$, $n_{14}=3154$, $n_{15}=973$,and $n_{16}=2220$.

The experimentally reconstructed ququart density matrix is, compared
with theoretical one, a bit more complex,
\begin{widetext}
\begin{equation}\label{mix1}
\rho_{exp}^{mix}=
    \left(
      \begin{array}{cccc}
        0.2300    &       0.0024+0.0009i       &       0.0211-0.0007i      &       -0.0015+0.0020i \\
        0.0024-0.0009i     &       0.0057      &       -0.0006+0.0017i      &      -0.0572-0.0019i \\
        0.0211+0.0007i     &       -0.0006-0.0017i      &       0.0041      &       0.0069+0.0018i \\
        -0.0015-0.0020i     &       -0.0572+0.0019i       &       0.0069-0.0018i      &       0.7571\\
      \end{array}
    \right).
\end{equation}
We obtain $Tr[\rho_{exp}^2]=0.634$ and $S = 0.394$ for the experimentally reconstructed ququart density matrix.
The fidelity is calculated to be $F=0.987$ and Fig.~\ref{mixed}(a) shows the graphical representation of the real part of the experimentally reconstructed mixed state ququart shown in eq.~(\ref{mix1}).

The experimentally reconstructed density matrix has the following inherent errors due to the fluctuations of the coincidence count rate.
\begin{equation}
\Delta\rho_{exp}^{mix}=
    \left(
      \begin{array}{cccc}
        0.0053          &       0.0027-0.0027 i       &       0.0030+0.0023i      &       0.0071+0.0044i \\
        0.0027+0.0027i    &      0.0008       &       0.0044-0.0070i      &       0.0044+0.0053i \\
        0.0030-0.0023i     &       0.0044+0.0070i       &      0.0007      &       0.0048-0.0048i \\
        0.0071-0.0044i     &      0.0044-0.0053i       &       0.0048+0.0048i      &       0.0096\\
      \end{array}
    \right).
\end{equation}
\end{widetext}

Let us now discuss the case in which the pump polarization is $45^\circ$. In this case, the two ququart basis states $|3\rangle$ and $|0\rangle$ are equally excited. Therefore, the theoretical ququart density matrix is given as,
\begin{equation}\label{mix0}
\rho_{theory}^{mix}=
    \left(
      \begin{array}{cccc}
        0.5000     &       0     &      0     &       0 \\
        0     &       0      &       0      &       0 \\
        0     &       0       &       0     &       0 \\
        0    &       0       &       0      &       0.5000\\
      \end{array}
    \right).
\end{equation}
The theoretical ququart density matrix results $Tr[\rho_{theory}^2]
= 0.5$ and $S = 0.5$. Note that for a complete mixture of all four
ququart basis states, $Tr[\rho^2]=1/4$ and $S = 1$.

For the equal mixture of $|3\rangle$ and $|0\rangle$, the tomographic coincidence measurement outcomes are (for $180$ s) $n_{1}=3442$, $n_{2}=30$, $n_{3}=3983$,
$n_{4}=23$, $n_{5}=1621$, $n_{6}=2358$, $n_{7}=1950$, $n_{8}=1895$,
$n_{9}=1906$, $n_{10}=1973$, $n_{11}=1959$, $n_{12}=1840$,
$n_{13}=2040$, $n_{14}=2026$, $n_{15}=1809$, and $n_{16}=1909$.

The experimentally reconstructed ququart density matrix for this
case is found to be,
\begin{widetext}
\begin{equation}\label{mix2}
\rho_{exp}^{mix}=
    \left(
      \begin{array}{cccc}
        0.4584     &       0.0142+0.0048i     &      0.0253+0.0162i     &       0.0158-0.0097i \\
        0.0142-0.0048i     &       0.0041      &       0.0012+0.0004i      &       -0.0406+0.0118i \\
        0.0253-0.0162i     &       0.0012-0.0004i       &       0.0031     &       0.0024+0.0029i \\
        0.0158+0.0097i    &       -0.0406-0.0118i       &       0.0024-0.0029i      &       0.5313\\
      \end{array}
    \right).
\end{equation}
The experimentally reconstructed ququart density matrix is characterized by $Tr[\rho_{exp}^2]=0.499$ and $S=0.504$. The fidelity is calculated to be $F=0.989$ and Fig.~\ref{mixed}(b) shows the graphical representation of the real part of the experimentally reconstructed mixed state ququart shown in eq.~(\ref{mix2}).

The error in the experimentally reconstructed density matrix, due to the count fluctuations, is found to be,
\begin{equation}
\Delta\rho_{exp}^{mix}=
    \left(
      \begin{array}{cccc}
        0.0078          &       0.0040-0.0039 i       &       0.0043+0.0036i      &       0.0075+0.0047i \\
        0.0040+0.0039i    &      0.0007       &       0.0047-0.0074i      &       0.0038+0.0047i \\
        0.0043-0.0036i     &       0.0047+0.0074i       &      0.0006      &       0.0042-0.0042i \\
        0.0075-0.0047i     &      0.0038-0.0047i       &       0.0042+0.0042i      &       0.0084\\
      \end{array}
    \right).
\end{equation}
\end{widetext}

\subsection{Partially-mixed state ququart}

A more general state of a ququart, as in eq.~(\ref{int}), can be prepared by transforming an initial mixed state of the form shown in eq.~(\ref{mix}). The same double-crystal scheme was used for generating the SPDC photon pair and, in this experiment, the pump polarization was $45^{\circ}$. A zero-order half-wave plate designed at 823.5 nm was used at the fast-axis angle of $22.5^{\circ}$ from the vertical axis.

\begin{figure}[t]
\centering\includegraphics[width=3.4in]{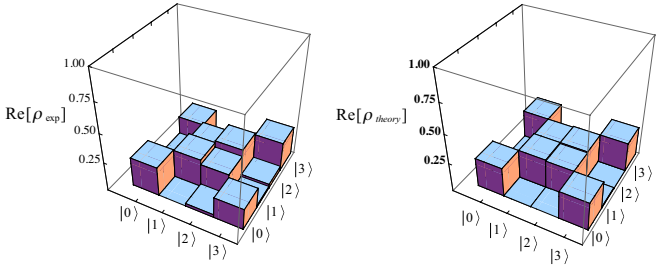}
\caption{A ququart with a partially-mixed state between a pure state and a completely mixed state. The calculated fidelity is $F=0.878$. We obtain $Tr[\rho_{exp}^2]=0.483$ and $S(\rho_{exp}) = 0.551$. }
\label{mid}
\end{figure}

The ququart density matrix calculated for this setting is,
\begin{widetext}
\begin{equation}
\rho_{theory}^{int}=
    \left(
      \begin{array}{cccc}
        0.2500     &       0       &       -.0086      &       0.2414+0.0644i \\
        0     &       0.2500       &       0.2414-0.0644i      &       0.0086 \\
        -0.0086     &       0.2414+0.0644i       &       0.2500      &      0 \\
        0.2414-0.0644i     &       0.0086       &       0     &       0.2500\\
      \end{array}
    \right).
\end{equation}
\end{widetext}
The above theoretical ququart density matrix results $Tr[\rho_{theory}^2]=0.5$ and $S=0.5$ which are equal to the values for the mixed state in eq.~(\ref{mix0}). This means that, even though the ququart state has been unitarily transformed to a new one with the help of a wave plate, the entropy of the system has not been changed. To actually reduce the entropy of the system, it is necessary to remove the temporal distinguishability between the amplitudes $|0\rangle$ and $|3\rangle$ which is introduced due to ultrafast pumping of the BBO crystals \cite{kim1}.  This can be accomplished, for example, by inserting a properly oriented quartz plate of the exact thickness in the path of the pump laser.

For the partially-mixed biphoton ququart state, the projection measurement resulted the following coincidence counts (for 180 s):  $n_{1}=1760$, $n_{2}=1730$, $n_{3}=1733$,
$n_{4}=1839$, $n_{5}=1687$, $n_{6}=1630$, $n_{7}=1758$,
$n_{8}=1961$, $n_{9}=817$, $n_{10}=3029$, $n_{11}=1008$,
$n_{12}=1701$, $n_{13}=1940$, $n_{14}=1944$, $n_{15}=1692$, and
$n_{16}=1192$.

The experimentally reconstructed density matrix is found to be,
\begin{widetext}
\begin{equation}
\rho_{exp}^{int}=
    \left(
      \begin{array}{cccc}
        0.2493     &       -0.0077-0.0007i       &       0.0234+0.0126i      &       0.1793+0.1665i \\
        -0.0077+0.0007i     &       0.2433       &       0.2032+0.1184i      &       0.0142-0.0036i \\
        0.0234-0.0126i     &       0.2032-0.1184i       &       0.2590      &       0.0333-0.0019i \\
        0.1793-0.1665i     &       0.0142+0.0036i       &       0.0333+0.0019i      &       0.2542\\
      \end{array}
    \right),
\end{equation}
which shows $Tr[\rho_{exp}^2]=0.483$ and $S=0.551$.

The inherent fluctuations of the count rate introduce the following error in the reconstructed density matrix, 
\begin{equation}
\Delta\rho_{exp}^{int}=
    \left(
      \begin{array}{cccc}
        0.0059          &       0.0042-0.0042 i       &       0.0046+0.0039i      &       0.0098+0.0036i \\
        0.0042+0.0042i    &      0.0059      &       0.0065-0.0054i      &       0.0043+0.0040i \\
        0.0046-0.0039i     &       0.0065+0.0054i       &      0.0061      &       0.0042-0.0042i \\
        0.0098-0.0036i     &      0.0043-0.0040i       &       0.0042+0.0042i      &       0.0059\\
      \end{array}
    \right).
\end{equation}
\end{widetext}

Figure \ref{mid} shows the real parts of the experimentally
reconstructed and the theoretical density matrices. The calculated
fidelity is somewhat low $F=0.878$ in this case.

\section{Discussion}

The fidelity $F$ quantifies how close the experimentally prepared
ququart state is to the one we intended to prepare, i.e., $F$ quantifies the overlap between the theoretical density matrix and the experimentally reconstructed density matrix. In section 4, we have analyzed errors introduced to the experimentally reconstructed density matrices due to the fluctuations of the count rates. As we have seen in eqs.~(8), (11), (14), and (17), however, these fluctuations contribute very small errors to the reconstructed density matrices. 

There are a number of external experimental factors which could strongly affect the fidelity. First, errors in the angular settings of the wave plates and polarizers used for the projection measurement. Significant improvement is possible by moving from hand-operated optic holders that are graduated in $1^\circ \sim
2^\circ$ increment to motorized holders. Second, less-than-ideal
spatial mode matching for the photon pairs coming from the two different
crystals. Mode matching can be implemented by adding a
spatial-filter or a short-piece of single-mode fiber and this should
improve the fidelity substantially. Third, inaccurate transformation matrices of the DBS. For the ququart tomography discussed in the previous section, accurate experimental reconstruction of the ququart density matrix requires the full knowledge of the polarization state change induced by all optical elements. In our experiment, the custom-made dichroic beam splitter (DBS) exhibited unexpected polarization-changing behaviors. It was found that the polarization states were changed both for the transmitted and the reflected beams. To account for the DBS behaviors, we carried out the Stokes parameter measurements for the transmitted and the reflected beams for six different input polarization states ($|H\rangle, |V\rangle, |45^\circ\rangle, |135^\circ\rangle, |R\rangle,$ and $|L\rangle$). From these measurements, it was possible to deduce the $2\times2$ DBS transformation matrices for the transmitted and the reflected modes. The experimental ququart density matrices shown in the previous section were reconstructed using the DBS matrices and therefore experimental errors introduced to the DBS matrices should have slightly affected the fidelity.

\begin{figure}[t]
\centering\includegraphics[width=3.0in]{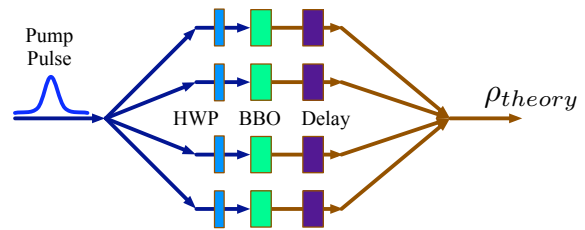}
\caption{Scheme for preparing a general biphoton ququart. Pump HWP
is used to control the relative magnitude of the SPDC amplitudes
generated at the BBO. Four BBO (two type-I and two type-II) crystals
are needed for this scheme. } \label{proposal}
\end{figure}

In our experiment, preparation of general states of a ququart was
demonstrated using the double-crystal scheme. This scheme, although
easy to setup, has potential difficulties in complete control of the
ququart states. For example, independent decoherence control for
amplitudes $|1\rangle$ ($|H_{\lambda_1},V_{\lambda_2}\rangle$) and
$|2\rangle$ ($|V_{\lambda_1},H_{\lambda_2}\rangle$) using a
birefringent medium is difficult as both amplitudes contain
horizontal and vertical polarization components.

For a complete control of the ququart state, i.e., to prepare a
ququart state with arbitrary values of $Tr[\rho^2]$ and $S$, we can
envision a four-crystal scheme in which a SPDC source is placed in
each arm of a four-path Mach-Zehnder interferometer as shown in
Fig.~\ref{proposal}. This scheme requires four crystals (two type-I
SPDC sources and two type-II SPDC sources) and the four-path
interferometer must be made stable for an accurate phase control.
Decoherence control can be accomplished by controlling the effective
beam paths of the interferometer arms using the tunable delays
installed at each of the beam path. (Note that arbitrary pure
ququart states can be prepared using just two crystals, see
Ref.~\cite{moreva1,bogdanov3}.)

It is, however, possible to design experimental schemes, for preparing arbitrary (mixed, partially-mixed, and pure) ququarts states, which are less complicated than the four-path interferometric scheme shown in Fig.~\ref{proposal}. In the following, we discuss two such experimental schemes, each of which are suited for particular ququart states in need.

\subsection{Scheme based on Mach-Zehnder interferometer}

The basic idea for the new ququart preparation scheme is based on the observation that the pure ququart state in eq.~(\ref{superposition}) can be re-written as,
\begin{widetext}
\begin{eqnarray}
|\Psi\rangle &=& c_{0}|H_{\lambda_1},H_{\lambda_2}\rangle + c_{1}|H_{\lambda_1},V_{\lambda_2}\rangle + c_{2} |V_{\lambda_1},H_{\lambda_2}\rangle + c_{3} |V_{\lambda_1},V_{\lambda_2}\rangle\nonumber \\
&=& |c_{0}| |H_{\lambda_1},H_{\lambda_2}\rangle +  |c_{3}|e^{i\phi_{03}} |V_{\lambda_1},V_{\lambda_2}\rangle + (|c_{1}| |H_{\lambda_1},V_{\lambda_2}\rangle+|c_{2}|e^{i\phi_{12}} |V_{\lambda_1},H_{\lambda_2}\rangle) e^{i\phi_{01}},
\label{eq:state}
\end{eqnarray}
\end{widetext}
where $\phi_{ij}$ ($i,j = 0,1,2,3$) is the relative phase for the $i$ and $j$ ququart basis states.  Here, we have put together the two ququart amplitudes that can be prepared with type-I SPDC (the first two terms) and the other two ququart amplitudes that can be prepared with type-II SPDC (the last two terms).

Experimentally, the first two terms in eq.~(\ref{eq:state}) can be prepared with two orthogonally oriented type-I BBO crystals placed in tandem as in Fig.~\ref{setup1}. The second two terms can then be prepared with two similarly placed type-II BBO crystals. Therefore, coherently or incoherently combining these two experimental schemes will allow us to prepare an arbitrary ququart states.

\begin{figure}[t]
\centering\includegraphics[width=3.4in]{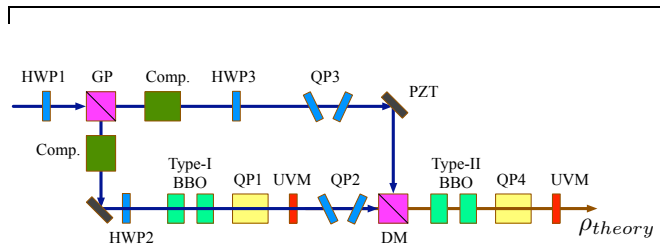}
\caption{Proposed setup for arbitrary ququart preparation. This scheme reduces the four-path interferometer shown in Fig.~\ref{proposal} to a Mach-Zehnder interferometer, making the experimental implementation much easier. QP2, QP3, and PZT adjust the phase terms in eq.~(\ref{eq:state}), $\phi_{03}$, $\phi_{12}$, and $\phi_{01}$, respectively.}
\label{setup_Shurupov}
\end{figure}

The experimental scheme realizing this idea is shown in Fig.~\ref{setup_Shurupov}. A Glan-Tompson prism (GP), transmitting the horizontally polarized component of the UV pump and reflecting the vertically polarized component, serves as the input beam splitter for the Mach-Zehnder interferometer. The reflected pump laser, after passing the birefringent compensator (Comp.) and a half-wave plate (HWP2), pumps a set of two orthogonally oriented type-I BBO crystals. A quartz plate (QP1) compensates the group velocity delay between the ordinary and the extra-ordinary polarized photons emitted from the pair of BBO crystals. The residual pump laser is removed by a uv mirror (UVM) and two quartz plates QP2 can be tilted along their optical axes to introduce a phase shift ($\phi_{03}$) between horizontally and vertically polarized type-I biphotons \cite{kim1}. The dichroic mirror (DM) is designed to transmit the biphoton wavelengths but to reflect the uv pump beam which comes from the upper path of the Mach-Zehnder interferometer.

The uv pump beam in the upper path of the Mach-Zehnder interferometer goes through a birefringent compensator, phase plates (QP3) for adjusting the phase ($\phi_{12}$) between horizontal and vertical components of the pump beam, and a piezoelectric translator (PZT) which introduces a relative phase shift ($\phi_{01}$) between the upper and the lower paths of the interferometer. The uv pump from the upper path, upon reflection at the DM, serves as the pump for the set of two orthogonally oriented type-II BBO crystals. The group velocity delay between biphoton amplitudes from the first and the second type-II BBO crystals are then compensated by the crystal compensator (QP4).

At the output of the experimental setup shown in Fig.~\ref{setup_Shurupov}, an arbitrary ququart states based on the biphoton polarization states of frequency-nondegenerate SPDC is prepared. We note that the scheme is loosely based on the biphoton qutrit setup demonstrated in Ref.~\cite{bogdanov1}, where three SPDC crystals are exploited for generation of  frequency-degenerate collinear biphotons.

\subsection{Scheme based on frequency non-degenerate and non-collinear regime of SPDC}

The interferometric scheme proposed in the previous section, although straightforward, might not be practical as the scheme inherits high phase sensitivity of the Mach-Zehnder interferometer which is not desirable for the purpose of preparing ququarts. Is it then possible to prepare an arbitrary ququart states using less complicated experimental scheme? The answer to this question is found to be positive, at least for the pure ququart states by using the biphoton polarization entangled states with controllable two-qubit concurrence $C$.

Let us first discuss the problem in a rather abstract form. In some arbitrary chosen four-dimensional computational basis, an arbitrary pure ququart state can be written as eq.~(\ref{superposition}). Since the biphoton ququart is in fact formed with two polarization qubits as defined in eq.~(\ref{basis}), we can make use of the fact that a unique set of orthonormal states of the two subsystems (two polarization qubits) $|A_i\rangle$ and $|B_i\rangle$ ($i=1,2$) exists such that the biphoton pure state in eq.~(\ref{superposition}) can be expressed in the form
\begin{equation}
|\psi\rangle = \sqrt{\chi_1} |A_1\rangle|A_2\rangle + \sqrt{\chi_2} |B_1\rangle|B_2\rangle.
\end{equation}
This is known as the Schmidt decomposition. The Schmidt coefficients $\chi_i$ are eigenvalues of the reduced density matrices of the subsystems. If a biphoton ququart state can be decomposed in this way (although the Schmidt coefficients and the Schmidt basis states may be different), one should be able to prepare an arbitrary pure ququart state, provided that the Schmidt coefficients and the Schmidt basis are experimentally controllable. The control of the Schmidt coefficients requires non-local unitary transformations which affect both subsystems but it is possible to use only local operations to switch between the Schmidt basis. For biphoton ququarts we discuss in this paper, these operations turn out to be rather simple as the subsystems are polarization qubits.

\begin{figure}[t]
\centering\includegraphics[width=3.4in]{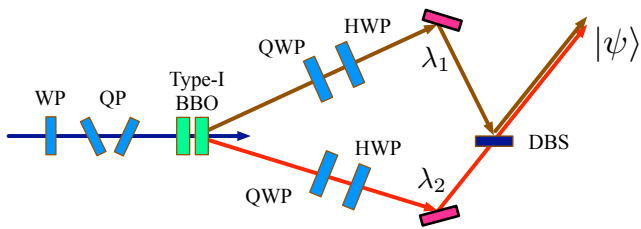}
\caption{Proposed experimental setup for preparation of an arbitrary pure ququart state. DBS represents the dichroic beam splitter. }
\label{setup_Straupe}
\end{figure}

The proposed setup to implement this idea is shown in Fig.\ref{setup_Straupe}. A set of two orthogonally oriented type-I BBO crystals, cut for frequency-nondegenerate non-collinear type-I SPDC, are pumped by a uv laser whose polarization is controlled by a half-wave plate (WP). A set of quartz plates (QP) introduces the relative phase between the horizontal and the vertical components of the uv pump. The initial polarization state of the biphoton generated in this process can be written as
\begin{equation}\label{last}
|\psi\rangle=\sqrt{\chi_1}|H_{\lambda_1}\rangle|H_{\lambda_2}\rangle+\sqrt{\chi_2}|V_{\lambda_1}\rangle|V_{\lambda_2}\rangle,
\end{equation}
where the relative magnitude of the coefficients is controlled by WP and the relative phase is controlled by tilting QP. For this biphoton state, the state of individual subsystem (qubit) is then given as,
\begin{equation}\label{rho2}
\rho_{i}=\chi_1|H_i\rangle\langle H_i|+ \chi_2|V_i\rangle\langle V_i|,
\end{equation}
where the subsystem index $i=1,2$.

So far we have produced a state with a given set of Schmidt coefficients (i.e., a given concurrence), but in a fixed (H-V) basis. The use of a quarter-wave plate (QWP) and and a half-wave plate (HWP) on each of the photons then locally and unitarily transforms the polarization states of the individual photon. Since the unitary transformation conserves the inner product, there always exists a state orthogonal to a given one in the two-dimensional Hilbert space. In other words, in eq.~(\ref{rho2}), if $|H_i\rangle\xrightarrow{U_i}|A_i\rangle$, then $|V_i\rangle\xrightarrow{U_i}|B_i\rangle$ with  $\langle A_i|B_i\rangle=0$. As a result, the state eq.~(\ref{last}) becomes
\begin{equation}\label{final}
|\psi\rangle\xrightarrow{U_1\otimes U_2}\sqrt{\chi_1}|A_1\rangle |A_2\rangle+\sqrt{\chi_2}|B_1\rangle|B_2\rangle,
\end{equation}
where $U_1$, for example, refers to the unitary polarization transformation for photon 1. Equation (\ref{final}) represents a general form of a pure biphoton polarization state or a pure biphoton ququart state.

The scheme shown in Fig.~\ref{setup_Straupe} is interferometrically more stable than the one in Fig.~\ref{setup_Shurupov} due to the fact that the photon pair from each type-I BBO crystal goes through the same optical paths. In principle, to prepare the state in eq.~(\ref{final}), the two paths in Fig.~\ref{setup_Straupe} should only be equal up to the coherence length of the pump laser. If the pump is broadband, however, it becomes necessary to further erase the temporal distinguishability between amplitudes from the first and the second BBO crystals, for example, by using a set of compensating crystals \cite{kim1}.

\section{Conclusion}

We have demonstrated that an arbitrary general single-ququart state can be prepared in a simple and controllable way by using the biphoton polarization state of ultrafast-pumped collinear frequency-nondegenerate SPDC. In addition, we have proposed two additional schemes which can be applied for arbitrary (pure and mixed) ququart state preparation.

Compared to other multi-dimensional quantum systems, the biphoton ququart is easier to prepare and characterize and states other than pure states can be prepared easily. Furthermore, it is possible to prepare a multi-ququart entangled state linear optically \cite{baek1,baek2}. We, therefore, believe that the general ququart state preparation scheme analyzed in this paper will find applications in quantum key distribution and quantum information processing.

\section*{Acknowledgments}

This work was supported, in part, by the Korea Research Foundation (R08-2004-000-10018-0 and KRF-2006-312-C00551), the Korea Science and Engineering Foundation (R01-2006-000-10354-0), Russian Foundation for Basic Research (Projects 06-02-16769 and 06-02-39015), the Leading Russian Scientific Schools (Project 4586.2006.2), and the Ministry of Commerce, Industry and Energy of Korea through the Industrial Technology Infrastructure Building Program.




\begin{thebibliography}{99}
\bibitem{disadvantage} Since the biphoton ququart is encoded at the two-photon polarization state of frequency-nondegenerate SPDC, fiber-based transmission of the biphoton ququart state is difficult due to dispersion. Free-space quantum communication protocols, therefore, are most likely candidates for potential applications.
\end{thebibliography}
\end{document}